\documentclass[twocolumn,aps,pr,superscriptaddress,preprintnumbers,nofootinbib,10pt]{revtex4-2}
\usepackage{amsmath}
\usepackage{amssymb}
\usepackage[dvipdf,dvips]{graphicx}
\usepackage{color}
\usepackage{hyperref}
\usepackage{url}
\usepackage{slashed}
\usepackage{subfigure}
\usepackage[usenames,dvipsnames]{xcolor}
\usepackage{amsfonts}
\usepackage{easyReview}
\usepackage{float}
\usepackage{amssymb}
\usepackage{epsfig}
\usepackage{graphics}
\usepackage{euscript}
\usepackage{slashed}
\usepackage{epstopdf}
\usepackage[utf8]{inputenc}
\allowdisplaybreaks
\usepackage[normalem]{ulem}
\usepackage{pifont}
\usepackage{dsfont}
\usepackage{MnSymbol}
\usepackage{verbatim}
\usepackage{graphicx}
\usepackage{latexsym}
\usepackage{tikz-feynman}
\usepackage[normalem]{ulem}

\begin{document}
	
	
	\title{Maximal violation of the Bell-CHSH inequality via bumpified Haar wavelets}
	\author{David Dudal} \affiliation{KU Leuven Campus Kortrijk–Kulak, Department of Physics,	Etienne Sabbelaan 53 bus 7657, 8500 Kortrijk, Belgium}\affiliation{Ghent University, Department of Physics and Astronomy, Krijgslaan 281-S9, 9000 Gent, Belgium}

	\author{Philipe De Fabritiis} \affiliation{CBPF --- Centro Brasileiro de Pesquisas Físicas, Rua Dr. Xavier Sigaud 150, 22290-180, Rio de Janeiro, Brazil}
	\author{Marcelo S.  Guimaraes} \affiliation{UERJ --- Universidade do Estado do Rio de Janeiro, Rua São Francisco Xavier 524, 20550-013, Rio de Janeiro, Brazil}
	\author{Itzhak Roditi} \affiliation{CBPF --- Centro Brasileiro de Pesquisas Físicas, Rua Dr. Xavier Sigaud 150, 22290-180, Rio de Janeiro, Brazil}
	\author{Silvio P. Sorella}	\affiliation{UERJ --- Universidade do Estado do Rio de Janeiro, Rua São Francisco Xavier 524, 20550-013, Rio de Janeiro, Brazil}


	\begin{abstract}
We devise a general setup to investigate the violation of the Bell-CHSH inequality in the vacuum state in the context of Quantum Field Theory. We test the method with massless spinor fields in $(1+1)$-dimensional Minkowski space-time.  Alice's and Bob's test functions are explicitly constructed, first by employing Haar wavelets which are then bumpified into proper test functions via a smoothening procedure relying on the Planck-taper window function. Relativistic causality is implemented by requiring the support of Alice's and Bob's test functions to be located in the left and right Rindler wedges, respectively. Violations of the Bell-CHSH inequality as close as desired to Tsirelson's bound are reported. We briefly comment on the extra portal, compared to earlier works, this opens to scrutinize Bell-CHSH inequalities with generic, interacting Quantum Field Theories.
	\end{abstract}

	\maketitle

\section{Introduction}
\vspace{-0.3cm}

Bell's inequalities have been a pivotal issue in Quantum Mechanics since their formulation~\cite{Bell64,Clauser69}. It is certainly appropriate to state that the principle of relativistic causality plays a key role in understanding the nature of this inequality. Requiring that Alice and Bob are space-like separated prevents any possible interference between their respective measurements, and it is worth reminding that the closure of the so-called causality loophole required highly sophisticated experimental tools \cite{Hensen15, Giustina15, Shalm15,Rosenfeld17, Li18, Storz23}. As far as relativistic causality is concerned, it seems natural to look at the Bell-CHSH inequality within the realm of Quantum Field Theory (QFT), a quite difficult endeavor, already investigated in the pioneering works~\cite{Summers87a,Summers87b,Summers87c} (see also \cite{Peruzzo22,Peruzzo23,Sorella:2023pzc,Dudal23,DeFabritiis23} for more recent attempts). In particular, the authors of \cite{Summers87a,Summers87b,Summers87c} have been able to show, by using methods of Algebraic QFT, that the Bell-CHSH inequality can be maximally violated already at the level of free fields.

\vspace{-0.3cm}

\section{Goal}
\vspace{-0.3cm}

The aim of the present work is to exhibit an explicit violation of the Bell-CHSH inequality in the vacuum within the QFT framework, thus giving continuity to the work done in \cite{Summers87a,Summers87b,Summers87c}. More precisely, we shall be able to provide a systematic construction of appropriate test functions needed to detect the Bell-CHSH inequality violation by means of wavelet representations. To our knowledge, this is the first time in which such an explicit construction is presented. We also highlight the difference between our approach and that of~\cite{Summers87a,Summers87b,Summers87c}.

\vspace{-0.3cm}

\section{Method}
\vspace{-0.3cm}

The whole procedure relies on the following:
\begin{itemize}
\vspace{-0.2cm}
\item[{\it (i)}] identify a Hermitian dichotomic field operator $\mathcal{A}$: ${\mathcal{A}}^{\dagger}= \mathcal{A}$ and $\mathcal{A}^2=1$.
\vspace{-0.2cm}
\item[{\it (ii)}] use smearing to localize the dichotomic field operators entering the Bell-CHSH inequality in suitable space-like separated regions of Minkowski space-time. Following  \cite{Summers87a,Summers87b,Summers87c}, we shall employ two pairs of smooth test functions with compact supports $(f,f') $ and $(g,g')$, referred to as Alice's and Bob's test functions, aka.~bump functions. Relativistic causality is implemented by demanding that the supports of $(f,f')$ and $(g,g')$ belong to Rindler's left and right wedges, respectively;
\vspace{-0.2cm}
\item[ {\it (iii)}] express the Bell-CHSH inequality in terms of the inner products between $(f,f')$ and $(g,g')$. The vacuum expectation value of the Bell-CHSH correlator in QFT,
\begin{equation}
\quad\;\;\; \langle \mathcal{C} \rangle = \langle 0 \vert \, i \left[\left( \mathcal{A}_f + \mathcal{A}_{f'}\right) \mathcal{A}_g + \left(\mathcal{A}_f - \mathcal{A}_{f'}\right) \mathcal{A}_{g'}\right] \vert 0 \rangle, \; \label{Bell-Corr}
\end{equation}
can then be reexpressed in terms of inner products between the test functions, namely,
\begin{align}\label{key-CB}
	\langle \mathcal{C} \rangle =  i \left( \langle f \vert g \rangle + \langle f \vert g' \rangle + \langle f' \vert g \rangle - \langle f' \vert g' \rangle \right),
\end{align}
where $\langle f \vert g \rangle $ stands for the Lorentz-invariant inner product, see Eq.~\eqref{INP1};
\vspace{-0.2cm}
\item[{\it (iv)}] show that  $\langle f \vert g \rangle$, $\langle f \vert g' \rangle$, $\langle f' \vert g \rangle$, $\langle f' \vert g' \rangle$ can be constructed so that the Bell-CHSH inequality is violated, {\it i.e.},
\begin{equation}
2 < | \langle \mathcal{C} \rangle | \le 2\sqrt{2}  \;, \label{violC}
\end{equation}
where the value $2\sqrt{2}$ is Tsirelson's bound \cite{Tsirelson80}. We shall proceed in two steps. First, we search for a preliminary set of would-be test functions $(\widetilde f, \widetilde f')$, $(\widetilde g, \widetilde g')$ adopting a Haar wavelet finite series representation. In the second step, the final form of the test functions $(f,f')$, $(g,g')$ is achieved via a bumpification procedure based on the Planck-taper window function. As a final result, we obtain explicit violations of the Bell-CHSH inequality as close as desired to Tsirelson's bound.
\end{itemize}

\vspace{-0.3cm}
\section{Outlook}
\vspace{-0.3cm}

Let us emphasize that nowadays there is a great interest in testing Bell-CHSH inequalities in high-energy physics~\cite{Fabbrichesi21, Severi22, Afik21, Afik22, Afik23, Barr22,Morales:2023gow}. This allows one to probe entanglement in an energy regime never explored before, in which the appropriate description of the physical phenomena is in the realm of QFT.  Our approach might, in principle, be used for any QFT, despite the numerical challenges that will naturally come with more complicated models, covering as well the case of interacting theories.  Let us limit here to mention that, in the interacting case, the inner products between the test functions are modified in such a way that the kernel corresponding to the free Wightman two-point function gets replaced by the K{\"a}llen-Lehmann spectral density, encoding the information about the interaction.

\vspace{-0.3cm}
\section{Spinor fields}
\vspace{-0.3cm}

	Let us introduce a QFT for a free spinor field in $(1+1)$-dimensions, with action
	\begin{align}
	S = \int \! d^2x \, \left[\bar{\psi} \left(i \gamma^\mu \partial_\mu - m \right) \psi \right].
	\end{align}
	In the above expression, $\psi_\alpha = \left(\psi_1, \psi_2\right)^t$ is a Dirac field described by a two-component spinor with complex $\psi_1, \psi_2$. In this work, we shall restrict ourselves to the free case and mainly consider the massless limit.
	The Clifford algebra is given by $\{\gamma^\mu , \gamma^\nu \} = 2 g^{\mu \nu}$, where the metric is $g_{\mu \nu} = \text{diag}(+1,-1)$ and the Dirac matrices are chosen as $\gamma^0 = \sigma_x$ and $\gamma^1 = i \sigma_y$, being $\sigma_x, \sigma_y$ the  Pauli matrices. According to  canonical quantization, we introduce the non-trivial equal-time anti-commutation relations
	\begin{align}\label{key10}
	\{ \psi_\alpha(t,x), \psi_\beta^\dagger(t,y)  \} = \delta_{\alpha \beta} \delta(x-y).
	\end{align}
	The Dirac field can be written in a plane wave expansion
	\begin{align}\label{key11}
	\psi(t,x) =\! \int \!\! \frac{dk}{2 \pi} \frac{m}{\omega_k} \left[ u(k) c_k \, e^{-i k_\mu x^\mu} + v(k) d_k^\dagger \, e^{+i k_\mu x^\mu}   \right],
	\end{align}
	where $k_\mu x^\mu = \omega_k t - kx$ and $\omega_k = \sqrt{k^2 + m^2}$.
	For the algebra of creation and annihilation operators, we get
	\begin{align}\label{key12}
		\{ c_k, c_q^\dagger  \} = \{ d_k, d_q^\dagger  \} = 2 \pi \frac{\omega_k}{m} \delta(k-q).
	\end{align}
	Evaluating the anti-commutators for different space-time points $x^\mu$ and $y^\mu$, we find
	\begin{align}\label{key13}
		\{ \psi_\alpha(x), \bar{\psi}_\beta(y)  \} = \left(i \gamma^\mu \partial_\mu - m\right)_{\alpha \beta} i \Delta_{PJ}(x-y),
	\end{align}
	with 
	\begin{align}\label{key14}
		i \Delta_{PJ}(x) = \int \!\! \frac{dk}{2 \pi} \frac{1}{2 \omega_k} \left(e^{-i k x} - e^{+i k x}\right).
	\end{align}
	The Pauli-Jordan distribution $\Delta_{PJ}$ is a real, Lorentz-invariant, odd under the exchange $x \rightarrow -x$, solution of the Klein-Gordon equation. Furthermore, this distribution vanishes outside the light cone ({\it i.e.}, $\Delta_{PJ}(x) = 0$ if $x^2 <0$),  ensuring that measurements at space-like separated points do not interfere (cf.~relativistic causality).
	
\vspace{-0.3cm}
\section{Smearing}
\vspace{-0.3cm}

	Quantum fields are operator-valued distributions~\cite{Haag:1992hx} and must be smeared in order to give well-defined operators acting on the Hilbert space. In the present case,  the smearing procedure is achieved by considering two-component spinor test functions of the form $h_\alpha (x) = \left(h_1(x), h_2(x)\right)^t$, where $h_1, h_2$ are commuting test functions belonging to the space $C_0^\infty (\mathbb{R}^4)$ of infinitely differentiable functions with compact support. For the  smeared spinor quantum fields we have
	\begin{align}\label{key15}
	\psi(h) &=  \int \! d^2 x \, \bar{h}^\alpha(x) \psi_\alpha(x), \nonumber \\
 \psi^\dagger(h) &=   \int \! d^2 x \, \bar{\psi}^\alpha(x) h_\alpha(x).
	\end{align}
	Due to the causal structure of the Pauli-Jordan distribution,  if we consider two test functions $(h, h')$ that have space-like separated supports, we will find $\{ \psi(h), \psi^\dagger(h') \} = 0$, which reflects causality at the level of smeared fields.
	
	From the definition of the smeared spinor field, by plugging the plane wave expansion, we find
	\begin{align}\label{key16}
		\psi(h) = c_h + d_h^\dagger, \quad \psi^\dagger(h) = c_h^\dagger + d_h,
	\end{align}
	where the smeared creation and annihilation operators read
	\begin{align}\label{key17}
	\!\!\!	c_h \! = \!\! \int \!\! \frac{dk}{2 \pi} \frac{m}{\omega_k} \bar{h}(k) u(k) c_k; \,\, d_h \! = \!\! \int \!\! \frac{dk}{2 \pi} \frac{m}{\omega_k} \bar{v}(k) h(-k) d_k,
	\end{align}
	with analogous equations for their conjugate expressions. From the canonical anti-commutation relations and the above definitions, one can compute the non-trivial anti-commutation relations in terms of the smeared creation and annihilation operators
	\begin{align}\label{key18}
		\{ c_h, c_{h'}^\dagger \} &= \int \! \frac{dk}{2 \pi} \frac{1}{2 \omega_k} \bar{h}(k) \left(\slashed{k} + m\right) h'(k), \nonumber \\
		\{ d_h, d_{h'}^\dagger \} &= \int \! \frac{dk}{2 \pi} \frac{1}{2 \omega_k} \bar{h'}(-k) \left(\slashed{k} - m\right) h(-k),
	\end{align}
	where in both expressions the constraint $\omega_k = \sqrt{k^2 + m^2}$ is implicitly understood, and we denote $\slashed{k} \equiv k_\mu \gamma^\mu $. 	
\vspace{-0.3cm}
\section{Bell setup}
\vspace{-0.3cm}

	Let us face now the introduction of a  Bell quantum field operator, Hermitian and dichotomic. Following \cite{Summers87a,Summers87b,Summers87c},  we shall consider the  smeared operator
	\begin{align}\label{key19}
		\mathcal{A}_h = \psi(h) + \psi^\dagger(h).
	\end{align}	
It is immediate to see that $	\mathcal{A}_h^\dagger = \mathcal{A}_h$. As it is customary, the inner product between test functions is obtained through the two-point Wightman function associated with the operator $\mathcal{A}_h$, that is, $ \langle h \vert h' \rangle  \equiv \langle 0 \vert \mathcal{A}_h \mathcal{A}_{h'} \vert 0 \rangle$. Using the anticommutation relations of the smeared creation and annihilation operators,  the vacuum expectation value of the product $\mathcal{A}_h \mathcal{A}_{h'}$ is easily evaluated, yielding
	\begin{align}\label{INP1}
	\langle h \vert h' \rangle = \!\! \int \!\!\! \frac{dk}{2 \pi} \frac{1}{2 \omega_k} &\big[\bar{h}(k) \left(\slashed{k} + m\right) h'(k) \nonumber \\
 &+ \bar{h'}(-k) \left(\slashed{k} - m\right) h(-k)\big].
	\end{align}
When  $h=h'$, we can identify the above expression as the norm squared of the test function  $h$. In particular, from the anti-commutation relations,
	\begin{equation}
	\langle \mathcal{A}_h^2 \rangle = \vert\vert h \vert\vert^2, \label{dich}
	\end{equation}
	showing that, as desired,  $\mathcal{A}_h	$ is a dichotomic operator, provided the test function $h$ is normalized to 1~\cite{Summers87a,Summers87b,Summers87c}. For  the Bell-CHSH correlator in the vacuum, we write
\begin{equation}
		\langle \mathcal{C} \rangle = \langle 0 \vert \, i \left[ \left( \mathcal{A}_f + \mathcal{A}_{f'}\right) \mathcal{A}_g + \left(\mathcal{A}_f - \mathcal{A}_{f'}\right) \mathcal{A}_{g'} \right] \vert 0 \rangle, \label{BCA1}
\end{equation}		
where, $(f,f')$ and $(g,g')$ are Alice's and Bob's test functions whose supports are located in Rindler's left and right wedges, respectively, and the factor $i$ is due to the anti-commuting nature of the spinor fields. The above expression can also be written in a Quantum Mechanics-like version:
\begin{equation}\label{key20}
	\langle \mathcal{C} \rangle =  i \left(\langle f \vert g \rangle + \langle f \vert g' \rangle + \langle f' \vert g \rangle - \langle f' \vert g' \rangle \right).
\end{equation}
As already stated, the main goal of this work is to explicitly construct test functions such that the Bell-CHSH inequality is violated. This amounts to find test functions $(f, f', g, g')$ belonging to the space $\mathcal{C}_0^\infty\left(\mathbb{R}^4\right)$, normalized to 1, such that $(f, f')$ and $(g, g')$ have space-like separated supports and, finally, such that we have $\vert \langle \mathcal{C} \rangle \vert > 2$. From the  definition, Eq.~\eqref{INP1}, imposing a reality condition on the test functions of the form $f_i^*(k) = f_i(-k), g_i^*(k) = g_i(-k)$, there is a simplification of the inner product expression, which becomes
	\begin{align}\label{key2}
		\langle f \vert g \rangle = \! \int \!\! \frac{dk}{2 \pi} \, &\Bigg[\left(\frac{\omega_k + k}{\omega_k}\right)f_1^*(k) g_1(k) \nonumber \\
  &+ \left(\frac{\omega_k - k}{\omega_k}\right)f_2^*(k) g_2(k) \Bigg].
	\end{align}
In particular, considering the massless case, we can rewrite the above expression  as
	\begin{align}\label{key3}
		\langle f \vert g \rangle = \int \! \! \frac{dk}{2 \pi} &\big[\left(1 + \text{sgn}(k)\right) f_1^*(k)g_1(k) \nonumber \\
		&+ \left(1 - \text{sgn}(k)\right) f_2^*(k)g_2(k) \big],
	\end{align}
where $\text{sgn}(k) = k / \vert k \vert$.  Going back to configuration space, we find 	$\langle f \vert g \rangle = I_1 + I_2$, where
	\begin{align}\label{key4}
		I_1 &=\! \int \! dx \, \left[f_1^*(x) g_1(x) + f_2^*(x) g_2(x)\right], \nonumber \\
		I_2 &=\! -\frac{i}{\pi} \! \int \!\!\! dx dy \left(\frac{1}{x-y}\right) \left[f_1^*(x) g_1(y) - f_2^*(x) g_2(y)\right].
	\end{align}
Here we will only consider real test functions. We remark that, with this assumption, we immediately find that $\langle f \vert f \rangle = I_1$, since the contribution $I_2$ vanishes by symmetry arguments under the exchange of $x$ and $y$. Furthermore, if $f$ and $g$ have disjoint supports, $\langle f \vert g \rangle = I_2\in i\mathbb{R}$.	

\vspace{-0.3cm}
\section{Wavelets}
\vspace{-0.3cm}

	Daubechies wavelets~\cite{Daubechies88, Daubechies92} are widely used to treat problems in signal processing and data compression~\cite{Kaiser94, Howard98}, and more recently, have been used in many different contexts, in QFT and beyond~\cite{Bulut13, George22, Haegeman18, Witteveen21, Deleersnyder21, Deleersnyder23}. The main idea here is to expand the test functions in terms of a finite number of Haar wavelets, a particularly useful type of Daubechies wavelets, well-known for their approximation abilities \cite{Haar}. These functions provide an orthonormal basis for the square-integrable functions on the real line and, moreover, have a compact support whose maximum size can be controlled. Let us introduce the mother wavelet $\psi$ as
 	\begin{align}  	
	\psi(x) = \left\{    \begin{array}{l l}
		&+1, \, {\rm if} \, x \in \big[ 0 , \frac{1}{2} \big),\\
		&-1, \, {\rm if} \, x \in \big[ \frac{1}{2} , 1 \big),\\
		&0, \, {\rm otherwise}.
	\end{array} \right.   	
	\end{align}
One  can then define the generic Haar wavelet $\psi_{n,k}$ as
	\begin{align}\label{key20}
		\psi_{n,k}(x) = 2^{n/2} \, \psi\left(2^n x -k\right)
	\end{align}
with support on $I_{n,k} = \big[ k 2^{-n}, (k+1) 2^{-n} \big)$ and piecewise constant, giving  $+ 2^{\frac{n}{2}}$ on the first half of $I_{n,k}$ and $- 2^{\frac{n}{2}}$ on the second half.  They satisfy
	\begin{align}\label{key21}
		\int \! dx \, \psi_{n,k}(x) \psi_{m,\ell}(x) = \delta_{nm} \delta_{k\ell}.
	\end{align}
Accordingly, for each would-be test function entering the Bell-CHSH inequality, we write 	
\begin{align}\label{key22}
		\widetilde f_j(x) = \sum_{n=n_i}^{n_f} \sum_{k=k_i}^{k_f} f_j(n, k) \, \psi_{n,k}(x),
	\end{align}
where $f_j(n, k)$ are the coefficients associated with the Haar wavelet basis element $\psi_{n,k}$ for the $j$-th component of the spinor test function $f(x)$. We remark that these parameters $\{n_i,n_f,k_i,k_f\}$ set the range and resolution of the Haar wavelet expansion.
	
In order to explicitly implement relativistic causality, we will consider the hypersurface $t=0$ and adopt the supports of $(f,f')$ corresponding to Alice's lab on the negative position axis, as well as the supports of $(g,g')$ corresponding to Bob's lab on the positive axis. This can be achieved taking $k \leq -1$ for $(f,f')$ and  $k\geq 0$ for $(g,g')$.

For the norm of the  test function $f$ we obtain
	\begin{align}\label{key23}
		\langle \widetilde f \vert \widetilde f \rangle = \sum_{n,k} \left[f_1^2(n,k) + f_2^2(n,k)\right]
	\end{align}
with analogous expressions for $\{\widetilde f', \widetilde g, \widetilde g'\}$. In the same vein, we can  evaluate $\langle \widetilde f \vert \widetilde g \rangle$, (naturally with similar expressions for $\langle \widetilde f' \vert \widetilde g \rangle$, $	\langle \widetilde f \vert \widetilde g' \rangle$ and $\langle \widetilde f' \vert \widetilde g' \rangle$), finding:

	\begin{align}\label{key24}
		\langle \widetilde f \vert \widetilde g \rangle &= \!\!\!\! \sum_{n,k,m,l} \!\!\!\! \left[f_1(n,k) g_1(m,l) - f_2(n,k) g_2(m,l)\right] \times \nonumber \\
		&\times \left[-\frac{i}{\pi} \int \!\! dx dy \, \left(\frac{1}{x-y}\right) \psi_{n,k}(x) \, \psi_{m,l}(y)\right].
	\end{align}
The advantage of using the Haar wavelet expansion is that all of the above integrals can be evaluated in closed form thanks to the piecewise constant nature of the wavelets. We refrain from listing the explicit expressions here as these are quite lengthy. Needless to say, numerical integration routines lead to consistent results. Therefore, given the parameters $\{n_i, n_f, k_i, k_f\}$, one can  obtain all the inner products, and then further manipulate the Bell-CHSH inequality, searching for the conditions to achieve its explicit violation. More precisely,  we shall impose
\begin{align}\label{SW}
\left\langle \widetilde f \vert \widetilde f\right\rangle & =  \left\langle \widetilde f' \vert \widetilde f'\right\rangle=\left\langle \widetilde g \vert \widetilde g\right\rangle=\left\langle \widetilde g' \vert \widetilde g'\right\rangle = 1 \,, \nonumber\\
\left\langle \widetilde f \vert \widetilde g\right\rangle &= \left\langle \widetilde f' \vert \widetilde g\right\rangle=\left\langle \widetilde f \vert \widetilde g'\right\rangle=-\left\langle \widetilde f' \vert \widetilde g'\right\rangle=-i \frac{\sqrt 2\lambda}{1+\lambda^2}
\end{align}
with $\lambda\in (\sqrt 2-1,1)$. It is then easily checked that $\left| \left \langle \mathcal{C} \right \rangle \right|= \frac{4\sqrt 2\lambda}{1+\lambda^2} \in (2,2\sqrt 2)$. Then, we can search for the wavelet coefficients that satisfy this constraint through a suitable numerical minimization procedure. Notice that the constraints \eqref{SW} are quadratic in nature, which in general lead to a well-posed problem, see e.g.~\cite{posed}.

\vspace{-0.3cm}
\section{Bumpification}
\vspace{-0.3cm}

	The strategy presented above still needs to be refined. The would-be test functions are not smooth, due to the jumps in the Haar wavelets. Nevertheless, there is a class of smooth bump functions with compact support, which can be used to approximate the Haar wavelets as precisely as desired.
	
Following \cite{McKechan:2010kp}, we define the basic Planck-taper window function with support on the interval $[0,1]$ by
	\begin{align}  	
	\sigma_0(x,\varepsilon) =  \left\{    \begin{array}{l l}
			&\left[1 + \exp\left(	\frac{\varepsilon  (2 x-\varepsilon )}{x (x-\varepsilon )}\right) \right]^{-1}, \, {\rm if} \, x \in \big( 0, \varepsilon \big),\\
			&+1, \, {\rm if} \, x \in \left[\varepsilon, 1-\varepsilon\right],\\
			&\left[1 + \exp\left(\frac{\varepsilon  (-2 x-\varepsilon +2)}{(x-1) (x+\varepsilon -1)}\right) \right]^{-1}, \, {\rm if} \, x \in \big( 1-\varepsilon, 1 \big),\\
			&0, \, {\rm otherwise}.
		\end{array} \right.   	
	\end{align}
	In the above expression, the parameter $\varepsilon$ regulates the fraction of the window over which the function smoothly rises from $0$ to $1$ and falls from $1$ to $0$. This gives a smooth version of the basic rectangle, the deviation with which can be made arbitrarily small by tuning $\varepsilon$.

\begin{figure}[t!]
	\begin{minipage}[b]{1.0\linewidth}
		\includegraphics[width=\textwidth]{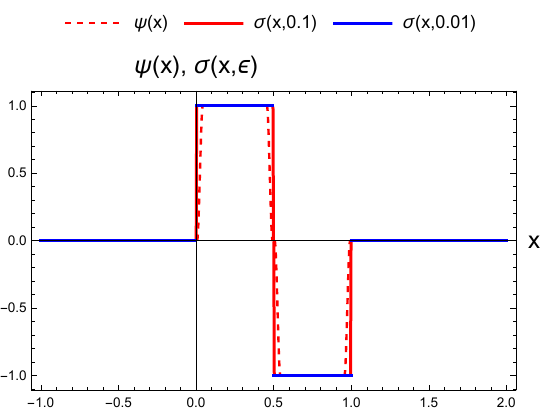}
	\end{minipage} \hfill
	\caption{The mother Haar wavelet and two of its bumpifications.}
	\label{fig1}
\end{figure}

With this object in hand, we then introduce the mother bump function with support on $[0,1]$ by
		\begin{align}  	
		\sigma(x,\varepsilon) =  \left\{    \begin{array}{l l}
		&+\sigma_0(2x, \varepsilon), \, {\rm if} \, x \in \big( 0, \frac{1}{2}\big),\\
		&-\sigma_0(2x-1,\varepsilon), \, {\rm if}\,  x \in \big( \frac{1}{2}, 1 \big),\\
		 &0, \, {\rm otherwise}.
		\end{array} \right.   	
	\end{align}
	Finally, we can define the $C_0^\infty(\mathbb{R})$ version of  the Haar wavelet,
	\begin{align}\label{key25}
		\sigma_{n,k}(x,\varepsilon) = 2^{n/2} \, \sigma(2^n x-k,\varepsilon).
	\end{align}
	This is indeed a smooth bump function with support on the interval $I_{n,k}$ that approximates as precisely as we want $\psi_{n,k}(x)$ per choice of $\varepsilon$, as illustrated in Fig.~\ref{fig1}. As such, each wavelet solution of the form \eqref{key22} can be replaced by a bumpified version,
\begin{align}\label{key22}
		f_j(x) = \sum_{n=n_i}^{n_f} \sum_{k=k_i}^{k_f} f_j(n, k) \, \sigma_{n,k}(x,\varepsilon),
	\end{align}
whilst retaining the various expansion coefficients $f_j(n, k)$, so that all crucial properties encoded in~\eqref{SW} are reproduced up to arbitrary precision if $\varepsilon$ is chosen small enough.

\begin{figure}[t!]
	\begin{minipage}[b]{1.0\linewidth}
		\includegraphics[width=\textwidth]{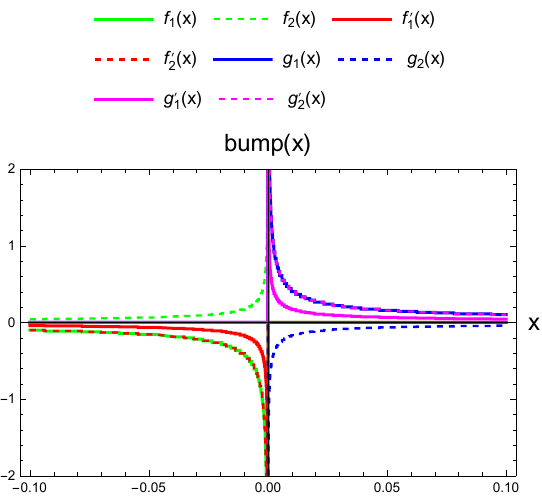}
	\end{minipage} \hfill
	\caption{Test function components for $\lambda=0.99$ with the following set of parameters: $\{n_i= -10; n_f = 120; k_i = -5; k_f = -1\}$ for $(f, f')$ and $\{m_i = -10; m_f = 120; \ell_i = 0; \ell_f = 4\}$ for $(g, g')$.}
	\label{fig2}
\end{figure}
\vspace{-0.3cm}
\section{Results}
\vspace{-0.3cm}

	Finally, we present and discuss our main results for the test functions leading to the violation of Bell-CHSH inequalities.  We will search for a solution by a numerical minimization (least squares fit so to say) of $\mathcal{R}=|\left\langle \widetilde f \vert \widetilde f\right\rangle-1|^2+|\left\langle \widetilde f'\vert \widetilde f'\right\rangle-1|^2+|\left\langle \widetilde g \vert \widetilde g\right\rangle-1|^2+|\left\langle \widetilde g' \vert \widetilde g'\right\rangle-1|^2+|\left\langle \widetilde f \vert \widetilde g\right\rangle+i \frac{\sqrt 2\lambda}{1+\lambda^2}|^2 + |\left\langle \widetilde f \vert \widetilde g'\right\rangle+i \frac{\sqrt 2\lambda}{1+\lambda^2}|^2 + |\left\langle \widetilde f' \vert \widetilde g\right\rangle+i \frac{\sqrt 2\lambda}{1+\lambda^2}|^2 + |\left\langle \widetilde f \vert \widetilde g'\right\rangle-i \frac{\sqrt 2\lambda}{1+\lambda^2}|^2$.  By choosing the wavelet basis sufficiently large, $\mathcal{R}$ becomes zero up to the desired precision, after which we stop the minimization. For the record, we also tested that directly minimizing $|\left\langle \widetilde f \vert \widetilde f\right\rangle-1|^2+|\left\langle \widetilde f' \vert \widetilde f'\right\rangle-1|^2+|\left\langle \widetilde g\vert \widetilde g\right\rangle-1|^2+|\left\langle \widetilde g' \vert \widetilde g'\right\rangle-1|^2+ |\left\langle \mathcal{C}\right\rangle - 4 \frac{\sqrt 2\lambda}{1+\lambda^2}|^2$ leads to the same solution.

              As a first test, we select $\lambda=0.7$, so that $\left|\langle \mathcal{C} \rangle\right| \approx 2.66$. To find the solution, we adopted the following parameter set $\{n_i= -5; n_f = 30; k_i = -4; k_f = -1\}$ for $(\widetilde f, \widetilde  f')$ and $\{m_i = -5; m_f = 30; \ell_i = 0; \ell_f = 3\}$ for $(\widetilde g, \widetilde g')$, with $\mathcal{R}=\mathcal{O}(10^{-26})$.  This already illustrates the effectiveness of our method and encourages us to search for larger violations. In order to do so, we need to correspondingly enlarge our wavelet basis, especially if we intend to approach Tsirelson's bound, $2 \sqrt{2}$, for $\lambda\to 1$. The larger basis allows not only for a larger covered sector of space, but in particular for a more pronounced peak in the test functions near the boundary of the causally disconnected intervals. Then the inner products between the causally disjoint test functions can get large enough towards an eventual saturation of the bound. As a matter of fact, imposing $\langle \mathcal{C} \rangle \approx 2.82$ for $\lambda=0.99$ and willing to achieve precision at the percent level, corresponding to $\mathcal{R}=\mathcal{O}(10^{-5})$, we were able to solve the constraints~\eqref{SW} if we adopt the following set of parameters: $\{n_i= -10; n_f = 120; k_i = -5; k_f = -1\}$ for $(f, f')$ and $\{m_i = -10; m_f = 120; \ell_i = 0; \ell_f = 4\}$ for $(g, g')$\footnote{Higher precision can be reached upon further enlarging the set of basis elements and, consequently, longer computation times. To get an idea about the computing time on a single standard laptop, solving the optimization presented here took a few minutes for the $\lambda = 0.7$ case with $1152$ coefficients and a few hours for the $\lambda = 0.99$ case with $5240$ coefficients. The wavelet coefficients for both reported cases can be obtained from the authors upon reasonable request.\vspace{0.3cm}}.

	For the bumpification, adopting the same set of wavelet coefficients, we thus replace the Haar wavelets $\psi_{n,k}(x)$ with  $\sigma_{n,k}(x)$. As a self-consistency test, we numerically computed the inner products again, now with these smooth functions expanded in terms of $\sigma_{n,k}(x)$, and checked if they are correctly normalized and violate the Bell-CHSH inequality up to the same precision as the underlying wavelet solution. For $\varepsilon = 10^{-10}$, we have found an excellent numerical agreement, as expected, showing that our strategy indeed works. The components of the corresponding normalized test functions leading to the Bell-CHSH inequality violation are shown in Fig.~\ref{fig2} for the case $\lambda=0.99$. It should be stressed that although the functions seem to increase without bounds near the origin, this is a misleading impression: all of them go to zero in the limit $x \rightarrow 0$, per construction. Also not visible in Fig.~\ref{fig2} is the fact that all shown functions do have compact support, again per construction.

Interestingly, there seem to be several reflection relations between the various test function components upon visual inspection of Fig.~\ref{fig2}. To verify these, we reconstructed the wavelet coefficients using an expansion with all the expected reflection relations built in, which resulted in a numerically indistinguishable solution from the one shown in Fig.~\ref{fig2}.

\vspace{-0.3cm}
\section{Comparison with earlier work of Summers--Werner}
\vspace{-0.3cm}

The seminal papers \cite{Summers87a,Summers87b,Summers87c} heavily relied on Tomita-Takesaki theory (see \cite{Witten:2018zxz} for an introduction to the latter) to prove the existence of a set of test functions so that the Tsirelson bound can be approximated as precisely as desired. To the best of our knowledge, the explicit form of the Summers-Werner test functions is, unfortunately, unknown. Notice that their construct is limited to the free field case, as Tomita-Takesaki theory has to date no interacting counterpart. From~\cite{Summers87a,Summers87b,Summers87c}, the test functions $(f,f') $ and $(g,g')$ are linear combinations of another set of causally disjoint test functions, $(f_1,f_2) $ and $(g_1,g_2)$, with certain constraints for the various inner products w.r.t.~Eq.~\eqref{key4}. It is an open question if the solution (strategy) proposed here leads to the same solution as the one of~\cite{Summers87a,Summers87b,Summers87c}; we will come back to this question in a larger forthcoming paper\footnote{As far as we know, there can be multiple sets of test functions leading to the same amount of Bell-CHSH violation. For instance, it is trivial to see that upon switching the sign of all upper or lower components of the spinor test functions does not change anything in the relevant inner products.}. As the number of test function constraints in~\cite{Summers87a,Summers87b,Summers87c} is larger than the ones we imposed --- in fact even more than necessary to attain the maximal violation which is due to the specific proof of~\cite{Summers87a,Summers87b,Summers87c} --- one might expect their and our solution will not be equivalent.

\vspace{-0.3cm}
\section{Conclusions}
\vspace{-0.3cm}

	We investigated the Bell-CHSH inequality in the context of QFT for a free massless spinor field in $1+1$ dimensions. Introducing suitable Bell operators built with smeared spinor fields, we defined an appropriate inner product associated with these operators through their Wightman functions. Expanding the would-be test functions used in the smearing procedure as a finite sum over Haar wavelets, we numerically constructed suitable coefficients leading to the violation of Bell-CHSH inequality, arbitrarily close to the maximal violation in fact. Using the Planck-taper window function, the discontinuous Haar wavelet solution set was then bumpified into $C_0^\infty(\mathbb{R})$ smooth functions with compact support up to arbitrary precision, allowing us to adopt the same set of wavelet coefficients that we found before. Therefore, we thus found a proper set of test functions leading to the explicit violation of the Bell-CHSH inequality in QFT. In future work, we foresee the generalization to the massive case, including to scalar field theories. Even more rewarding will be to test the here presented bumpified wavelet method for interacting QFTs, in which case far less is known about the possibility of having maximal violation or not. An interacting $(1+1)$-dimensional fermionic theory like the Thirring model~\cite{Thirring:1958in} will constitute the most interesting test bed, especially since the spectral function is known exactly \cite{Johnson:1961cs,Thompson:1983yr}, and the latter will enter the inner product as we already alluded to in the main text. 
 Finally, we plan to apply the method presented here to field theories in 1+2 and 1+3 dimensions, covering also the case of Bose fields by means of Weyl operators, see~\cite{Weyl}.
 We will report on these and other matters in future work.

{\it Acknowledgments ---}
	The authors would like to thank the Brazilian agencies CNPq and FAPERJ for financial support. S.P.~Sorella, I.~Roditi, and M.S.~Guimaraes are CNPq researchers under contracts 301030/2019-7, 311876/2021-8, and 310049/2020-2, respectively. PDF is grateful to Gustavo P. de Brito, Henrique S. Lima, and Letícia F. Palhares for interesting discussions, and to Pedro C. Malta for helpful comments on the draft.


\end{document}